\documentclass[11pt]{article}

\usepackage{epsfig}
\newcommand{\beq}{\begin{equation}}
\newcommand{\eeq}{\end{equation}}
\newcommand{\beqa}{\begin{eqnarray}}
\newcommand{\eeqa}{\end{eqnarray}}
\newcommand{\beqar}{\begin{eqnarray*}}
\newcommand{\eeqar}{\end{eqnarray*}}

\newcommand{\be}{\begin{equation}}
\newcommand{\ee}{\end{equation}}
\newcommand{\bea}{\begin{eqnarray}}
\newcommand{\eea}{\end{eqnarray}}
\newcommand{\ba}{\begin{array}}
\newcommand{\ea}{\end{array}}

\def\bbox{{\,\lower0.9pt\vbox{\hrule \hbox{\vrule height 0.2 cm
\hskip 0.2 cm \vrule height 0.2 cm}\hrule}\,}}
\newcommand{\dsl}{\pa \kern-0.5em /}

\newcommand{\M}{{\cal M}}
\newcommand{\N}{{\cal N}}

\newcommand{\cijk}{C_{IJK}}
\newcommand{\hone}{h_{(1,1)}}

\newcommand{\nperp}{\nabla_\perp^2}

\begin{document}


\thispagestyle{empty}
\vspace{40pt}

\hfill{hep-th/0305261}

\vspace{128pt}

\begin{center}
\textbf{\Large From wrapped M-branes to Calabi-Yau}\\
\vspace{18pt}
\textbf{\Large black holes and strings}
\vspace{40pt}

David Kastor

\vspace{12pt} \textit{Department of Physics}\\
\textit{University of Massachusetts}\\
\textit{Amherst, MA 01003}\\
\vspace{6pt}
\texttt{kastor@physics.umass.edu}
\end{center}

\vspace{40pt}

\begin{abstract}

We study a class of $D=11$ BPS spacetimes that describe M-branes wrapping supersymmetric $2$ and $4$-cycles of Calabi-Yau $3$-folds.  We analyze the geometrical significance of the supersymmetry constraints and gauge field equations of motion for these spacetimes.  We show that the dimensional reduction to $D=5$ yields known BPS black hole and black string solutions of $D=5$, $N=2$ supergravity. The usual ansatz for the dimensional reduction is valid only in the linearized regime of slowly varying moduli and small gauge field strengths.   Our identification of the massless $D=5$ modes with $D=11$ quantities extends beyond this regime and should prove useful in constructing non-linear ansatze for Calabi-Yau dimensional reductions of supergravity theories.

\end{abstract}

\setcounter{footnote}{0}

\newpage

\setcounter{equation}{0}

\section{Introduction}

Branes wrapping supersymmetric cycles of string/M-theory compactifications have been studied in many different contexts in recent years, using a variety of different methods.  In this paper we will consider supergravity solutions for such brane configurations that take into account the full higher dimensional spacetime structure.  In particular, we will study BPS solutions of $D=11$ supergravity describing $M$-branes wrapping supersymmetric cycles of Calabi-Yau manifolds.  

The general forms of these spacetimes, which we will call FS spacetimes, and the constraints imposed on them by supersymmetry have been analyzed in the series of papers \cite{Fayyazuddin:1999zu}\cite{Gomberoff:1999ps}\cite{Fayyazuddin:2000em}\cite{Brinne:2000fh}\cite{Cho:2000hg}
\cite{Brinne:2000nf}\cite{Husain:2002tk}\cite{Husain:2003df}\cite{Husain:2003ag}.  However, solving the combined system of supersymmetry constraints and gauge field equations of motion to obtain the actual spacetime fields for a specified brane configuration has proven to be quite difficult.  Thusfar, explicit solutions have been found only in the near horizon regime \cite{Fayyazuddin:2000em}\cite{Maldacena:2000mw}\cite{Gauntlett:2000ng}\cite{Acharya:2000mu}
\cite{Gauntlett:2001qs}\cite{Gauntlett:2001jj}.  In this limit, the global structure of the Calabi-Yau manifold drops out.

In this paper, we will study FS spacetimes in the opposing limiting case, the far field limit.  Here, we will see, the topology of the Calabi-Yau manifold and the wrapped cycle dominate the structure of the solutions.  In this limit, the solutions may be viewed as BPS black branes in the dimensionally reduced spacetime.  We will study the examples of $M2$-branes and $M5$-branes wrapping respectively supersymmetric $2$-cycles and $4$-cycles of Calabi-Yau three-folds.  These correspond to BPS black holes and black strings of $D=5$, $N=2$ supergravity.  We compare the FS spacetimes in this limit with the $D=5$ black holes of reference \cite{Sabra:1998yd} and black strings of reference \cite{Chamseddine:1998yv}.  We will see that this comparison yields considerable insight into the system of equations governing FS spacetimes.

Additional motivation for this work comes from the following questions.  The $D=5$ solutions of references \cite{Sabra:1998yd}\cite{Chamseddine:1998yv} depend only on the homology class of the wrapped brane.  One can ask whether $D=11$ spacetimes exist that depend on the detailed position of the cycle within the Calabi-Yau space, or is some sort of smearing of sources mandatory?  Similarly, it should be interesting to see how the attractor mechanism \cite{Ferrara:1995ih} is manifest in the full higher dimensional metric.

We shall also see below that the comparison of the $D=11$ FS spacetime fields with the corresponding $D=5$ fields is complicated by the lack of a consistent truncation scheme.  In the $AdS_4\times S^7$ compactification of $D=11$ supergravity to gauged $D=4$ supergravity, for example, explicit nonlinear expressions for the $D=11$ fields in terms of the massless $D=4$ fields have been given in references \cite{Nastase:1999cb}\cite{Nastase:1999kf}.   Solutions to the $D=4$ equations of motion may be lifted to solutions to the $D=11$ equations of motion via these expressions.  In contrast, the ansatze for dimensional reduction on Calabi-Yau manifolds, of $D=10$, or $D=11$ supergravities \cite{Ferrara:1988ff}\cite{Bodner:1990zm}\cite{Ferrara:ik}\cite{Cadavid:1995bk}\cite{Bohm:1999uk}
are valid only to linear order in variations of the moduli.  The detailed form of the correspondence between $D=5$ and exact $D=11$ solutions should provide useful clues towards a nonlinear form of the ansatz.  One interesting feature of the $D=11$, FS spacetimes is that the metric on the internal Calabi-Yau space is not Kahler, so that even the proper definition of the moduli fields needs more careful treatment.

\section{$D=11$ Supergravity on Calabi-Yau Three-folds}\label{reduce}

We begin by briefly recalling the dimensional reduction of $D=11$ supergravity on a Calabi-Yau three-fold $\M$ to $D=5$, $N=2$ supergravity \cite{Cadavid:1995bk}.  Here we use the term Calabi-Yau manifold to mean a compact, Kahler manifold with vanishing first Chern class.  In the present section, we additionally assume that the metric on $\M$ is Ricci-flat and Kahler.  However, in subsequent sections this will not be the case.

Ricci-flat, Kahler metrics on $\M$ are parameterized by $h_{(1,1)}$ Kahler moduli and $h_{(2,1)}$ complex structure moduli \cite{Candelas:1990pi}.  After dimensional reduction, the corresponding scalar fields become components of $h_{(1,1)}-1$ vector multiplets and $h_{(2,1)}+1$ hypermultiplets.  The 
vector multiplet scalars correspond to the Kahler moduli subject to the constraint of fixed volume for $\M$.  
The overall volume modulus becomes part of the universal hypermultiplet.  The hypermultiplets will not be relevant to our discussion of $M$-branes wrapping Kahler calibrated cycles, and we will ignore them in the following.  They would be important for an analogous discussion of $M$-branes wrapping SLAG calibrated cycles. 

The bosonic fields of $D=11$ supergravity are the metric $G_{MN}$\footnote{The conventions we use for coordinate indices are as follows:  $M,N=0,1,\dots,10$ are $D=11$ coordinate indices, 
$m,n=1,2,3$ are complex coordinate indices on $\M$, and $\mu,\nu=0,1,2,3,4$ are indices for the $D=5$ non-compact coordinates.}
and $3$-form gauge potential $A_{MNP}$.  The action for the bosonic fields is given by
\be
S= {1\over 2 k_{11}^2}\left (\int d^{11}x \sqrt{-G}\Big( R[G]-{1\over
48}F^{MNPQ}F_{MNPQ}\big) -{1\over 6} \int A\wedge F\wedge F \right ).
\ee
The dimensional reduction proceeds by taking the Ricci-flat metric on $\M$ to depend on the $D=5$ coordinates through the Kahler moduli scalars $\Phi^I$, $I=1,\dots,h_{(1,1)}$.  The Kahler $(1,1)$-form $J$ is given in terms of the Hermitian metric $g_{m\bar n}$ on $\M$ by $J=ig_{m\bar n}dz^m dz^{\bar n}$.  If $\alpha^I$ are a basis of homology $2$-cycles, then the moduli are given by 
$ \Phi^I=\int_{\alpha^I} J $.
Small variations in the metric on $\M$ that preserve the Ricci-flatness condition and leave the complex structure fixed can then be expressed \cite{Candelas:1990pi} in terms of small variations of the moduli fields by
\be
\delta g_{m\bar n}= -i \omega_{I\, m\bar n}\delta\Phi^I,
\ee
where $\omega_I$ form a harmonic basis for $H^{(1,1)}(\M,Z)$ with respect to $g_{m\bar n}$ normalized such that
$\int_{\alpha^I}\omega_J=\delta^I_J$. 
Within cohomology we then have the equality
\be\label{cohomology}
J = \omega_I\Phi^I.
\ee
A total of $\hone$ massless $D=5$, $1$-form gauge fields $A^I_\mu$
arise from taking the $D=11$, $3$-form gauge potential to have the form
\be\label{gaugeansatz}
A_{\mu m\bar n}= A^I_\mu\ \omega_{Im\bar n}.
\ee
One linear combination of these vector fields becomes the graviphoton, part of the graviton
multiplet, leaving $\hone -1$ vector fields, which combine with the constrained Kahler moduli to form the bosonic content of the $\hone -1$ vector multiplets.

The bosonic action of $D=5$, $N=2$ supergravity that results from the dimensional reduction is then given by
\begin{equation}\label{action}
S = {1\over 2\kappa_5^2}\int d^5x \sqrt{-g}\left\{ R -{1\over 2} G_{IJ} \partial_{\mu} \Phi^I \partial^\mu \Phi^J
- {1\over 2} G_{IJ} F_{\mu\nu} {}^I
F^{\mu\nu J}
+{1\over 24} \epsilon^{\mu\nu\rho\sigma\lambda} C_{IJK} 
F_{\mu\nu}^IF_{\rho\sigma}^JA_\lambda^K \right\}.
\end{equation}
Here $C_{IJK}$ are a set of integer constants given by the integral
\be\label{intersection}
C_{IJK}=\int_\M \omega_I\wedge\omega_J\wedge\omega_K,
\ee
which is well-defined on cohomology.  This computation can also be recast in terms of the intersection number of homology $4$-cycles.  In this formulation the integer-valuedness of $C_{IJK}$ is immediately apparent.  The volume form on $\M$ is given by $J\wedge J\wedge J/6$.  By virtue of equations (\ref{cohomology}) and (\ref{intersection}) the overall volume of $\M$ is then given by 
$ V(\Phi)={1\over 6} \cijk \Phi^I\Phi^J\Phi^K$.  The  
vector multiplet scalars are constrained to leave the overall volume of $\M$ constant.  If we fix this constant to unity, then the constraint on the scalars $\Phi^I$ in the action (\ref{action}) becomes
\be\label{volumeconstraint}
{1\over 6} \cijk \Phi^I\Phi^J\Phi^K =1.
\ee
%
The moduli space metric, which also serves as the gauge coupling matrix in (\ref{action}), is given by
\be\label{modulimetric}
G_{IJ}=-{1\over 2} \partial_I\partial_J \ln V(\Phi) |_{V=1}.
\ee
Finally we note that an alternate formulation of the dimensional reduction could be given in terms of dual moduli associated with the integrals of $J\wedge J$ over a basis of homology $4$-cycles.  Such a reformulation turns out to be useful for the reduction of wrapped M2-branes to $D=5$ black holes and will be discussed below in section (\ref{blackholes}).

\section{M5-branes wrapping 4-cycles and $D=5$ Black Strings}\label{strings}

An M5-brane wrapping a supersymmetric $4$-cycle of a Calabi-Yau 3-fold yields a
BPS black string in $5$-dimensions.  The resulting $5$-dimensional solutions have
been given in reference \cite{Chamseddine:1998yv}.  We want to see how these
black string spacetimes emerge as appropriate limits of the $11$-dimensional FS
spacetimes for M5-branes wrapping $4$-cycles that were given in reference
\cite{Cho:2000hg}.

\subsection{Black strings} 
The $D=5$ black string spacetimes of reference 
\cite{Chamseddine:1998yv} are given by
\bea\label{blackstring}
ds^2 = f^{-1/3}(-dt^2+dy^2) + f^{+2/3}\delta_{\alpha\beta}dx^\alpha dx^\beta \\
\Phi^I= f^{-1/3} f^I \label{moduli}\\
F^I_{\alpha\beta}=\epsilon_{\alpha\beta\gamma}\delta^{ \gamma\rho}\partial_\rho
f^I . \label{bsgauge}
\eea
The string lies along the $y$ axis and $x^\alpha$ with $\alpha=1,2,3$ are the spatial directions transverse to the string.  The functions $f^I(x^\alpha)$ are harmonic with respect to the flat metric on the
$3$-dimensional transverse space, {\it i.e.} $\nabla_\perp^2f^I=0$ with
$\nabla^2_\perp=\delta^{\alpha\beta}\partial_\alpha\partial_\beta$.  Finally, the metric
function $f$ is determined by the relation
\be
f={1\over 6}C_{IJK}f^If^Jf^K 
\ee
which implements the constraint (\ref{volumeconstraint}) that the volume of the
internal Calabi-Yau space is constant.

\subsection{Wrapped M5-branes}\label{m5section}
The $D=11$, FS spacetimes for M5-branes
wrapping $4$-cycles of Calabi-Yau 3-folds \cite{Cho:2000hg} are given by 
\be\label{eleven}
ds^2=   H^{-1/3}(-dt^2+dy^2)+H^{2/3} \delta_{\alpha\beta}dx^\alpha dx^\beta +
2 H^{-1/3} k_{m\bar n}dz^m dz^{\bar n},
\ee
\be\label{elevengauge}
A_{tz m\bar n r\bar s}= H^{-1} ( k_{m\bar n}k_{r\bar s}-k_{m\bar s}k_{r\bar
n}).
\ee
where $z^m$ with $m=1,2,3$ are complex coordinates on the Calabi-Yau space $\M$.  
The function $H$ and the Hermitian metric $k_{m\bar n}$ depend both on position on $\M$ and on the noncompact transverse coordinates.
Note that the Hermitian metric $g_{m\bar n}$ on $\M$ is written as a conformal rescaling $g_{m\bar n}= H^{-1/3}k_{m\bar n}$ of the Hermitian metric $k_{m\bar n}$.  This rescaling is convenient because it simplifies the constraints imposed by supersymmetry on the spacetime fields.  In particular, if we consider the metric on $\M$ with fixed transverse coordinates $x^\alpha$, then supersymmetry requires that $k_{m\bar n}$ be a Kahler metric on $\M$ \cite{Husain:2003df}\footnote{In reference \cite{Cho:2000hg} it was simply assumed that the metric $k_{m\bar n}$ in 
(\ref{eleven}) was Kahler.  In reference \cite{Husain:2003df} it was shown that this requirement is imposed by supersymmetry in the present case.  However, for similarly constructed wrapped M2-brane spacetimes the condition imposed by supersymmetry turns out in some cases to be less restrictive \cite{Brinne:2000nf}\cite{Husain:2002tk}\cite{Husain:2003df}.  One instance of this will be discussed in the following section.}
This implies that the unrescaled metric 
$g_{m\bar n}= H^{-1/3}k_{m\bar n}$ will also be Kahler only if the function $H$ is  constant on $\M$, {\it i.e.} if $H$ depends only on the transverse coordinates.

Supersymmetry also requires that the function $H$ and the Kahler metric $k_{m\bar n}$
satisfy the further constraints 
\be\label{constraintone}
\partial_\alpha\log (k/H)=0
\ee
\be\label{constrainttwo}
\partial_m\partial_{\bar n}\log (k/H)=0,
\ee
where $k=\det(k_{m\bar n})$.  We can now note that, since the complex dimension of $\M$ is $3$, the determinant of the unrescaled metric is given by $g=\det g_{m\bar n}=k/H$, which is the quantity that appears in (\ref{constraintone}) and (\ref{constrainttwo}).  These constraints then have the following simple geometrical
interpretations.  Equation (\ref{constraintone}) implies that the volume element
on $\M$ is independent of the transverse coordinates. 
In particular, this implies that the total volume of $\M$ is constant.  This is in accordance with the overall
volume constraint (\ref{volumeconstraint}).

Equation (\ref{constrainttwo}) is a Ricci-flatness condition on $\M$, but not for the usual Levi-Civita connection. 
Recall that a complex manifold $\M$ with a 
Hermitian metric $g_{m\bar n}$, admits a unique, metric compatible, Hermitian
connection given in components by $\Gamma_{mn}^p=g^{p \bar r}\partial_m g_{n\bar r}$ and 
$\Gamma_{\bar m\bar n}^{\bar p}=g^{\bar p  r}\partial_{\bar m} g_{\bar n r} $.
The Ricci tensor determined by this Hermitian connection has nonzero components given by
\be\label{ricci}
R_{m\bar n}=-\partial_m\partial_{\bar n}\log g.
\ee
We then recognize the left hand side of (\ref{constrainttwo}) as minus the Ricci
tensor of the Hermitian connection compatible with the unrescaled metric
$g_{m\bar n}$ on $\M$.  Finally, note that the Hermitian connection will
be torsion free and hence coincide with the Levi-Civita connection only if the metric
$g_{m\bar n}$ is Kahler.  We have noted above that this will not generally be the case.  
Hence, equation (\ref{constrainttwo}) does not generally imply that $g_{m\bar n}$ is Ricci-flat with respect
to the Levi-Civita connection.  This makes sense, of course, because the gauge field stress-energy tensor has non-zero components in the directions tangent to $\M$.  It seems interesting that despite this a simple, alternative Ricci-flatness condition holds\footnote{The vanishing of the first Chern class of $\M$ implies that the Ricci form associated with (\ref{ricci}) must be trivial in cohomology.  The condition $R_{m\bar n}=0$ is certainly consistent with this.}.

Finally, we examine the equations of motion for the $7$-form gauge field strength.  
After making use of the constraints imposed by supersymmetry, only the equation 
$\nabla_A F^{Atym\bar n r\bar s}=0$ yields a new constraint.  If we define the rescaled Kahler $(1,1)$-form
$J_k=i k_{m\bar n}dz^m dz^{\bar n}$ and consider the function $H$ to be a $(0,0)$-form on $\M$, 
then this component of the gauge field equations takes the simple form
\be\label{motion}
\nperp J_k + 2\partial\bar\partial H=0 .
\ee
This says that acting with the transverse Laplacian on the Kahler form $J_k$ yields an exact form on $\M$.

\subsection{Dimensionally reducing the wrapped M5-brane spacetimes} 
We now want to see that the $D=5$ black string spacetimes (\ref{blackstring}) are indeed the dimensional reductions
of the $D=11$ wrapped M5-brane spacetimes given in (\ref{eleven}) and (\ref{elevengauge}).  We could a priori carry out this comparison in two different ways.  We could try to lift the $D=5$ solutions to $D=11$, or we could try to reduce the $D=11$ solutions to $D=5$.  

The lifting method is hampered by the lack of a nonlinear ansatz for $D=11$ fields in terms of the massless $D=5$ degrees of freedom.  Ideally, given a configuration of the massless $D=5$ fields, which solve the $D=5$ equations of motion, we would have a formula that lifts it to a configuration of $D=11$ fields that solve the $D=11$ equations of motion.  Such a nonlinear ansatz has recently been found for the $AdS_4\times S^7$ compactification of
$11$-dimensional supergravity \cite{Nastase:1999cb}
\cite{Nastase:1999kf}.  At present, however, for Calabi-Yau compactifications we can only do this with very limited accuracy, {\it i.e.} to linear order in small variations of the moduli and small $D=5$ gauge field strengths.  This approximation is sufficient for deriving the dimensional reduction of $D=11$ supergravity to $D=5$, $N=2$ supergravity, as sketched in section (\ref{reduce}).  However, it ignores, for example, the backreaction of the gauge field stress-energy, which would enter at quadratic order, on the Calabi-Yau metric.  If we do not ignore these terms, then the Calabi-Yau metric will not be Ricci-flat.

We could restrict our comparison to the region in which the linearized approximation is good, so that the metric on $\M$ is Ricci-flat.  In the $D=5$ spacetimes, it is straightforward to identify this region. The linearized approximation becomes increasingly good in the far field limit, {\it i.e.} far from the position of the brane in the transverse space, where the field strengths and derivatives of the moduli fields are small.  However, $D=5$, $N=2$ supergravity governs the dynamics of the massless $D=5$ fields exactly.  Moreover in the near filed region of the $D=5$ black string spacetimes, derivatives of the moduli fields become large as do the gauge field strengths.  Therefore, we would like to be able to compare the $D=5$ and $D=11$ spacetimes throughout the transverse space.  

Our strategy will be to start by making the comparison in the far field region, where the linearized approximation is valid and the metric on $\M$ is approximately Ricci-flat.  In this region, it is straightforward to extract the massless $D=5$ field content from the $D=11$ fields.  Based on these results, we will see that there is a natural extension of this map from $D=11$ fields to massless $D=5$ fields that holds throughout the transverse space.   

In order to begin, we need to identify the region of the $D=11$ FS spacetimes in which the linearized approximation holds and the metric on $\M$ becomes approximately Ricci-flat.  From the analysis of the supersymmetry constraints in section (\ref{m5section}) we know that the metric $g_{m\bar n}$ on $\M$ becomes Ricci-flat in the limit that the function $H$ is constant on $\M$.  This must therefore be the case in the far field region of interest.  Hence, in making our comparison we start by looking at the FS spacetimes with the assumption that $H$ is constant on $\M$.
Examining the FS spacetime metric (\ref{eleven}), we see that the $D=5$ part of the metric matches the black string metric (\ref{blackstring}) if we simply identify $H=f$.
The moduli of the Calabi-Yau part of the $D=11$ metic are given in this limit by $\Phi^I=H^{-1/3}h^I$, where the $h^I$ are moduli of the rescaled Kahler form $J_k$, 
\be
h^I(x^\alpha)=\int_{\alpha^I} J_k.
\ee
Making the identification $h^I=f^I$, we see that the moduli of the FS spacetimes will have the same form as the moduli scalars of the black string spacetime (\ref{moduli}), if we can show that the functions $h^I$ are harmonic in the transverse space.
Turning to the gauge field equation of motion (\ref{motion}), we see that the second term vanishes since $H$ is constant along $\M$ in the far field limit.  Integrating equation (\ref{motion}) over a homology $2$-cycle $\alpha^I$, then gives
\be\label{harmonic}
\nperp h^I=0
\ee
and we see the functions $h^I$ are indeed harmonic in the transverse space.  We have not yet checked agreement for the gauge fields.  We will see below that this works as well.  Subject to this result, we have now seen that the massless fields extracted from the $D=11$ FS spacetimes agree with those of the $D=5$ black string in the far field, linearized regime where the Calabi-Yau metric is approximately Ricci-flat.

We now want to extend the identification of massless $D=5$ modes with quantities from $D=11$ throughout the transvere space.  We start by recalling that the metric $g_{m \bar n}$ on $\M$ will not in general be Kahler.  The associated Kahler form will consequently not be closed, and its integrals over $2$-cycles will not be invariant under deformations within a given homology class.  Therefore, moduli cannot be defined with respect to $g_{m \bar n}$ outside of the linearized approximation.  The conformally rescaled metric $k_{m\bar n}$, however, is Kahler and its moduli $h^I$ defined above continue to be well defined even outside the linearized regime.  Additionally, the $h^I$ continue to be harmonic functions throughout the transverse space, {\it i.e.} equation (\ref{harmonic}) continues to hold outside the linearized regime.  The second term in (\ref{motion}) no longer vanishes.  However, it is an exact form and its integral over a homology cycle $\alpha^I$ necessarilly vanishes.  Hence the moduli of the rescaled metric $k_{m\bar n}$ are harmonic functions even close to the brane in the transverse space where the $D=5$ fields are varying rapidly and the linearized approximaion fails.  The identification $h^I=f^I$ between the harmonic functions can then be extended throughout the transverse space.  

We now construct a $D=5$ function $h(x^\alpha)$ from the $D=11$ function $H(z^m,z^{\bar n},x^\alpha)$ 
by integrating over $\M$ in the following way
\be\label{volume}
h^{-1} = {1\over V_k}\ \int_{\M} H^{-1}\    d(vol)_k,
\ee
where $d(vol)_k$ and $V_k$ are the volume element and the total volume of $\M$ with respect to the conformally rescaled metric $k_{m\bar n}$.  This extracts the harmonic component of $H^{-1}$ with respect to $k_{m\bar n}$. Since the volume form associated with the Kahler metric $k_{m\bar n}$ is given by ${1\over 6}J_k^{\ 3}$, we have the result
$V_k = {1\over 6}\ C_{IJK}\, h^I\, h^J\, h^k$.
Because the volume $V$ of the Calabi-Yau manifold $\M$ with respect to the unrescaled metric $g_{m \bar n}$ is simply the right hand side of (\ref{volume}) without the $1/V_k$ prefactor, we then have the relation
$V= {1\over 6}\ h^{-1} C_{IJK}\, h^I\, h^J\, h^k$, together with the constraint (\ref{constraintone}) that $V$ is constant in the transverse space.  We can then normalize the volume to one, $V=1$, and identify the function $h$ with the metric function $f$ in the black string spacetime (\ref{blackstring}).

Finally, we must see how the gauge field strength of the wrapped M5-brane in (\ref{elevengauge}) dimensionally reduces to that of the $D=5$ black string in (\ref{blackstring}).  To understand the correct way to proceed, note that if we regard an M5-brane as an electric object with $7$-form field strength, then its charge is computed by integrating the dual $4$-form field strength over a $4$-surface $\N$ that surrounds the brane.  Similarly, the charge of a $D=5$ black string, regarded as an electric object, is obtained by integrating the $2$-form dual of its field strength over an $S^2$ that surrounds it.  We reduce from $D=11$ to $D=5$ by decomposing the $4$-surface surrounding the M5-brane as $\N=\Sigma\times S^2$, where $\Sigma$ is a $2$-surface in $\M$ and the $S^2$ surrounds the $D=5$ black string.  Only the homology class of $\Sigma$ within $\M$ is important, so all the information about the $M5$-brane charge is contained in the integrals of the dual $4$-form field strength over the basis homology cycles $\alpha^I$.  Making use of equation (\ref{constraintone}) the $(1,1)$-form part of the dual field strength reduces to
\be
F_{\alpha\beta m\bar n}= i\, \epsilon_{\alpha\beta\gamma}\,\delta^{\gamma\lambda}\, \partial_\lambda k_{m\bar n}.
\ee
Integration over the basis $2$-cycles $\alpha^I$ then correctly gives the $D=5$, $2$-form field strengths
\begin{eqnarray}
F^I_{\alpha\beta} & = &\int_{\alpha^I} F_{\alpha\beta m\bar n}\ dz^mdz^{\bar n} \\
& = &\epsilon_{\alpha\beta\gamma}\, \delta^{\gamma\lambda}\, \partial_\lambda \int_{\alpha^I} J_k \\
& = &\epsilon_{\alpha\beta\gamma}\, \delta^{\gamma\lambda}\, \partial_\lambda h^I
\end{eqnarray}
which match those of the black string spacetime as given in equation (\ref{bsgauge}).  This completes the identification of the black string spacetimes as the dimensional reduction of the $D=11$ wrapped M5-brane spacetimes.
We have seen, in particular, that although the metric on the Calabi-Yau space is not Kahler, that an alternative set of moduli parameters can be definied in terms of the rescaled metric $k_{m\bar n}$ and that these coincide with the harmonic functions appearing in the $D=5$ massless scalars $\Phi^I$.

\section{M2-branes wrapping 2-cycles and $D=5$ Black Holes}\label{blackholes}

We now turn to M2-branes wrapping supersymmetric $2$-cycles, which dimensionally reduce these to $D=5$ black holes.  Because many of the important points arose already in section (\ref{strings}), the discussion in this section will be brief.  However, we will see that there is an interesting interplay between the role played by dual moduli in the black hole solutions and the fact that supersymmetry of the $D=11$, FS spacetimes requires only that the square of the Kahler form $J_k$ be a closed form \cite{Husain:2002tk}.

\subsection{Black holes} 
The $D=5$ black hole spacetimes \cite{Sabra:1998yd} are given by 
\bea
ds_5^2=-f^{-4/3} dt^2 +f^{2/3}\delta_{\alpha\beta}dx^\alpha dx^\beta \label{bhmetric}\\
\Phi_I={1\over 3} f^{-2/3}f_I \label{bhmoduli}\\
G_{IJ}F^J_{0\alpha}={1\over 2}f^{-4/3}\partial_\alpha f_I,\label{bhgauge}
\eea
where $\alpha,\beta=1,2,3,4$, $G_{IJ}$ is the moduli space metric (\ref{modulimetric}) and the functions $f_I$ satisfy $\nperp f_I=0$.  Here the fields $\Phi_I$ are dual moduli that implicitly determine the moduli $\Phi^I$ through the relation
\be\label{dual}
\Phi_I={1\over 6}C_{IJK}\Phi^J\Phi^K.
\ee
The metric function $f$ is determined by the constraint (\ref{volumeconstraint}) that the overall volume modulus be constant, which can also be rewritten as the condition
$ \Phi_I\Phi^I=1$.

The dual moduli determine the cohomology class of the square of the Kahler form on $\M$.  By Poincare duality, there exists a basis of homology $4$-cycles $\beta_I$, such that the oriented intersection numbers $\#(\cdot,\cdot)$ of the $\beta_I$ with the homology $2$-cycles $\alpha^J$ are given by
$ \#(\alpha^I,\beta_J)=\delta^I_J$.
We can now choose a basis $\lambda^I$ for $H^4(M,Z)$ such that $\int_{\beta_J}\lambda^I=\delta^I_J$.  We then also have the relation
$\int_\M \omega_J\wedge\lambda^I =\delta^I_J$.
The dual moduli defined by
\be
\Phi_I={1\over 6} \int_{\beta_I} J\wedge J
\ee
are then related to the moduli $\Phi^I$ by equation (\ref{dual}).

The appearance of the moduli space metric $G_{IJ}$ in the expression for the electric field of the black hole (\ref{bhgauge}) is another indication of the role played by $4$-cycles on $\M$ in these solutions.  One can formulate $D=11$ supergravity in terms of the dual $7$-form field strength and its $6$-form gauge potential.  Massless $D=5$ gauge fields then corresponding to taking 
\be 
A_{\mu\nu k\bar l m\bar n} = A_{I\, \mu\nu}\tilde\omega^I_{\ k\bar l m\bar n}
\ee
where the $\tilde\omega^I$ are a basis for harmonic $(2,2)$ forms normalized such that $\int_{\beta_J}\tilde\omega^I=\delta^I_J$ and related to the basis of harmonic $(1,1)$-forms by
${}^*\omega_I=G_{IJ}\tilde\omega^J$.  The $D=5$, $3$-form field strengths corresponding to (\ref{bhgauge})  are then given simply in terms of the functions $f_I$ by
\be\label{dualbhgauge}
F_{I\, \alpha\beta\gamma}={1\over 2}\epsilon_{\alpha\beta\gamma\rho}\delta^{\rho\sigma}\partial_\sigma f_I.
\ee
This form of the gauge field facilitates comparison with the $D=11$ spacetimes.

\subsection{Wrapped $M2$-branes} 
The $D=11$, FS spacetimes for M2-branes wrapping 2-cycles of Calabi-Yau three-folds \cite{Cho:2000hg} are given by 
\be\label{FStwo}
ds_{11}^2= -H^{-4/3}dt^2+H^{2/3} d\vec x\cdot d\vec x +2 H^{-1/3} k_{m\bar n}dz^m
dz^{\bar n},\quad
A_{tm\bar n}= iH^{-1} k_{m\bar n}
\ee
As shown in reference \cite{Brinne:2000nf}\cite{Husain:2002tk}, the rescaled Kahler form $J_k$ in this case need not be closed.  Supersymmetry of (\ref{FStwo}) requires only the weaker condition
\be\label{closed}
dJ_k^{\ 2}=0,
\ee
where $J_k^{\ 2}\equiv J_k\wedge J_k$.
The further constraints of supersymmetry are again given by equations (\ref{constraintone}) and (\ref{constrainttwo}), with the same geometrical interpretation discussed above.
The gauge field equations of motion reduce to a single equation on $(2,2)$-forms\footnote{This equation is given incorrectly in reference \cite{Cho:2000hg}.  The correct gauge field equation for an M2-brane wrapping a supersymmetric $2$-cycle of a Calabi-Yau $N$-fold, for $N=2,3,4,5$ is given by
$\nabla_\perp^2J_k^{\ N-1}+2i(N-1)\partial\bar\partial(H J_k^{\ N-2})=0.$  It was found in \cite{Husain:2002tk} that
supersymmetry requires that the Kahler form $J_k$ satisfy $dJ_k^{\ N-1}=0$.  Therefore dual moduli defined by integrals of $J_k^{\ N-1}$ would be harmonic in the transverse space.}
\be\label{m2equation}
\nabla_\perp^2 J_k^{\ 2}+4i\partial\bar\partial(H J_k)=0.
\ee
We see in this case that the transverse Laplacian acting on $J_k^{\ 2}$ gives an exact form on $\M$.

\subsection{Dimensionally reducing the wrapped M2-brane spacetimes} 

The condition $dJ_k^{\ 2}=0$ dovetails nicely with the role played by dual moduli in the $D=5$ solutions.  
As in section (\ref{strings}), in this example we can only define moduli associated with the Kahler form $J$ in the far-field limit, where the Calabi-Yau metric becomes Ricci-flat and Kahler.  Since $dJ_k\ne 0$, we also cannot define moduli associated with the rescaled Kahler form.  However, we can define dual moduli associated with the closed form $J_k^{\ 2}$ according to  
\be
h_I={1\over 6}\int_{\beta_I}J_k^{\ 2}.
\ee
Integrating equation (\ref{m2equation}) over a homology $4$-cycle $\beta_I$ then gives $\nperp h_I=0$ and we can make the identification $h_I=f_I/3$ throughout the transverse space.  Again we can define a $D=5$ function $h$ from the $D=11$ function $H$ by equation (\ref{volume}).  In the far field limit, $h=H$ and the dual moduli of $\M$ and the $D=5$ metric then reproduce equations (\ref{bhmoduli}) and (\ref{bhmetric}) if we set $f=h$.  It is also then straightforward to check in this limit that the $D=11$ gauge field reduces to (\ref{dualbhgauge}).

\section{Conclusions}

We have shown that the $D=11$, FS spacetimes of reference \cite{Cho:2000hg} together with the conditions on teh Kahler form found in reference \cite{Husain:2002tk} dimensionally reduce to $D=5$ Calabi-Yau black hole and black strings.  This lets us conclude that the $D=11$ spacetimes do indeed describe wrapped brane configurations.  We are able to identify the $D=5$ massless fields with properties of the $D=11$ spacetimes throughout the transverse space.  This includes regions in which the moduli vary rapidly and field strengths are large, in which the linearized approximation used to construct the dimensionally reduced theory breaks down.  Further scrutiny of these results should provide useful clues towards constructing non-linear ansatze for Calabi-Yau reductions of $D=11$ supergravity.  One could start by asking, for example, what class of metrics on the Calabi-Yau space are relevant for the dimensional reduction.  
This is similar to questions asked in recent discussions of compactifications with flux, {\it e.g.} reference \cite{Gurrieri:2002wz} in which the condition $dJ^{\ 2}=0$ on the Kahler form also arises.

In sections (\ref{strings}) and (\ref{blackholes}) we have discussed the geometrical significance of the supersymmetry constraints and gauge field equations of motion for FS spacetimes.  This analysis should be useful in making progress on other questions raised in the introduction.   It should also be helpful in organizing the search for further examples of FS spacetimes describing M-branes wrapping SLAG calibrated cycles in Calabi-Yau spaces, or supersymmetric cycles in spaces of exceptional holonomy.

\section{Acknowledgements}

I thank Hyunji Cho, Moataz Emam and Jennie Traschen for there helpful input on this subject.  This work was supported in part by National Science Foundation grant PHY-9801875.


\end{document}